\documentclass[prl,superscriptaddress,twocolumn,showkeys,preprintnumbers,floats,floatfix]{revtex4}

\usepackage{graphicx}
\usepackage{amsmath}

\begin{document}

\title{Dynamically controlled charge sensing of a few-electron silicon quantum dot}

\author{C. H. Yang}
\affiliation{Centre of Excellence for Quantum Computation and Communication Technology, School of Electrical Engineering \& Telecommunications, The University of New South Wales, Sydney 2052, Australia}
\author{W. H. Lim}
\affiliation{Centre of Excellence for Quantum Computation and Communication Technology, School of Electrical Engineering \& Telecommunications, The University of New South Wales, Sydney 2052, Australia}
\author{F. A. Zwanenburg}
\affiliation{Centre of Excellence for Quantum Computation and Communication Technology, School of Electrical Engineering \& Telecommunications, The University of New South Wales, Sydney 2052, Australia}
\author{A. S. Dzurak}
\affiliation{Centre of Excellence for Quantum Computation and Communication Technology, School of Electrical Engineering \& Telecommunications, The University of New South Wales, Sydney 2052, Australia}

\date{\today}

\begin{abstract}
We report charge sensing measurements of a silicon metal-oxide-semiconductor quantum dot using a single-electron transistor as a charge sensor with dynamic feedback control.
Using digitally-controlled feedback, the sensor exhibits sensitive and robust detection of the charge state of the quantum dot,
even in the presence of charge drifts and random charge rearrangements.
The sensor enables the occupancy of the quantum dot to be probed down to the single electron level.

\end{abstract}



\maketitle

Non-invasive charge sensing~\cite{Field1993} is an invaluable tool for the study of electron charge and spin states in nanostructured devices. It has been used to identify electron occupancy
down to the single electron level ~\cite{Elzerman2003,Simmons2007} and has made possible the single-shot readout of
single electron spins confined in both quantum dots~\cite{Elzerman2004} and dopants~\cite{Morello2010}.
Both quantum point contacts (QPCs) and single electron transistors (SETs) possess high transconductance,
making them sensitive to their local electrostatic environment and therefore excellent charge sensors.
A QPC can be conveniently integrated with lateral quantum dot structures
formed in two-dimensional electron layers in GaAs/AlGaAs ~\cite{DiCarlo2004,Petta2004,Johnson2005,Cassidy2007},
Si/SiGe~\cite{Simmons2007,Simmons2009} and Si metal-oxide-semiconductor field-effect transistors (MOSFETs)~\cite{Nordberg2009,Xiao2010}.
SETs also have been integrated with quantum dots in a variety of structures including
Ge/Si core/shell nanowires~\cite{Hu2007}, carbon nanotubes~\cite{Biercuk2006}, graphene~\cite{Wang2010}
and Si MOSFETs~\cite{Podd2010}.

In this Letter we demonstrate a silicon MOS quantum dot~\cite{Angus2007,Lim2009b}
integrated with a nearby SET charge sensor~\cite{Angus2008},
using a dynamic feedback technique to maintain constant charge sensitivity over a wide operating range.
The feedback algorithm dynamically adjusts the gate-voltage of the SET to ensure that it operates at a constant output current.
In this configuration the SET optimally responds to changes in the local electrostatic environment,
such as electron tunnelling events. In contrast with previous sensing measurements,
the dynamic feedback employed here allows the SET sensor to recover from random charging events
that would otherwise reduce the sensitivity of the sensor.
This enables high-sensitivity measurements to be obtained over many hours,
even in the presence of random and large charge rearrangements.

Figure 1 shows a scanning electron microscope (SEM) image and schematic of the structure used for these experiments.
The structure was fabricated using a multi-layer Al-Al$_2$O$_3$-Al gate stack~\cite{Lim2009b}
on a high resistivity ($\rho$ $>$ 10~k$\Omega$cm at 300~K) silicon substrate with SiO$_2$ gate oxide of thickness 10~nm.
The fabrication process is similar to that described in Lim \textit{et al.}~\cite{Lim2009}.
In this structure, the roles of the quantum dot and SET charge sensor are interchangeable.
Here, we operate the lower device as the quantum dot while the upper device acts as the charge sensor.

\begin{figure}[t]
\includegraphics[width=8.5cm]{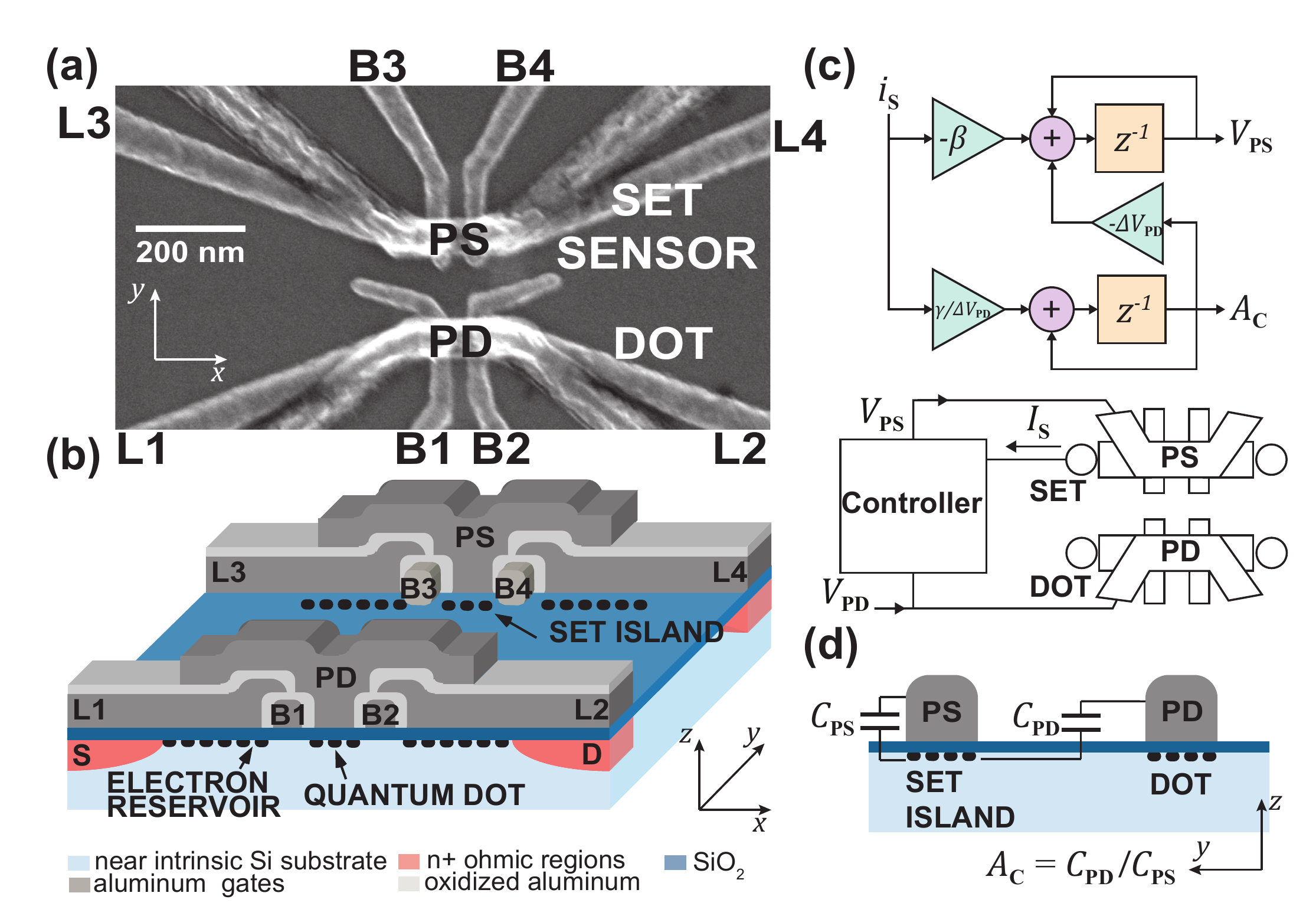}
\caption{(a) SEM image of the silicon MOS device, the bottom one operating as the dot and the upper one as the charge sensor.
		(b) The 3D schematic model of the device.
		(c) Block diagram of the second-order feedback control system used to ensure
		stable and continuous charge sensing of the quantum dot by the SET sensor.
		(d) Physical interpretation of mutual capacitance ratio, $A_\textrm{C}=C_\textrm{PD}/C_\textrm{PS}$.}
\label{fig1}
\end{figure}

Ten gate electrodes are used to electrostatically define the two devices,
which can be independently measured.
Positive voltages on the gates are used to induce electron accumulation layers below the Si-SiO$_2$ interface.
Barrier gates B1-B4 produce tunnel barriers in the electron layers, as shown in Fig. 1(b).
In both devices the dot (or SET island) electron occupancy is controlled by a ``plunger" gate,
labelled PD for the dot and PS for the SET,
while ``lead gates" L1-L4 are used to induce the four source and drain electron reservoirs for the two devices.

The electrical measurements were performed in a dilution refrigerator at a base temperature of $\sim${}40 mK.
We operated the quantum dot with an ac excitation voltage $V_\textrm{sd}$ of 100 $\mu$eV at 87 Hz
and the SET with $V_\textrm{sd}$ of 400 $\mu$eV at 133 Hz.
The lead gates of both dot and SET were fixed at $V_\textrm{L1,L2,L3,L4}$ = 3.0~V
to induce the electron reservoirs.
The barrier gates were operated in the range $V_\textrm{B1,B2,B3,B4}$ = 0.6--0.7~V.
Further details on the operation of such multi-gate MOS nanostructures are given in Lim \textit{et al.}~\cite{Lim2009b}.

The spacing between the quantum dot and the island of the SET charge sensor was $\sim${}120~nm. We also studied devices separated by 1 micron, where an additional metallic antenna was used to
enhance the capacitive coupling from the dot to the SET island~\cite{Hu2007, Biercuk2006, Buehler2006}.
In those devices the electron accumulation layer in the leads strongly screened the capacitance from antenna to both islands,
preventing effective charge sensing.
We therefore found that a directly coupled device, as shown in Fig. 1, was the most effective.

Charge sensing of the quantum dot by the SET is performed by measuring both devices simultaneously, as depicted in Fig. 1(c).
Using the SET plunger gate PS we tune the current $I_\textrm{S}$ of the SET to the edge of a Coulomb peak,
where the transconductance $dI_\textrm{S}$/$dV_\textrm{PD}$ is high, thus enabling the detection of single-electron transfers in the quantum dot, the latter controlled using its plunger gate PD.

\begin{figure}[t]
\includegraphics[width=8.5cm]{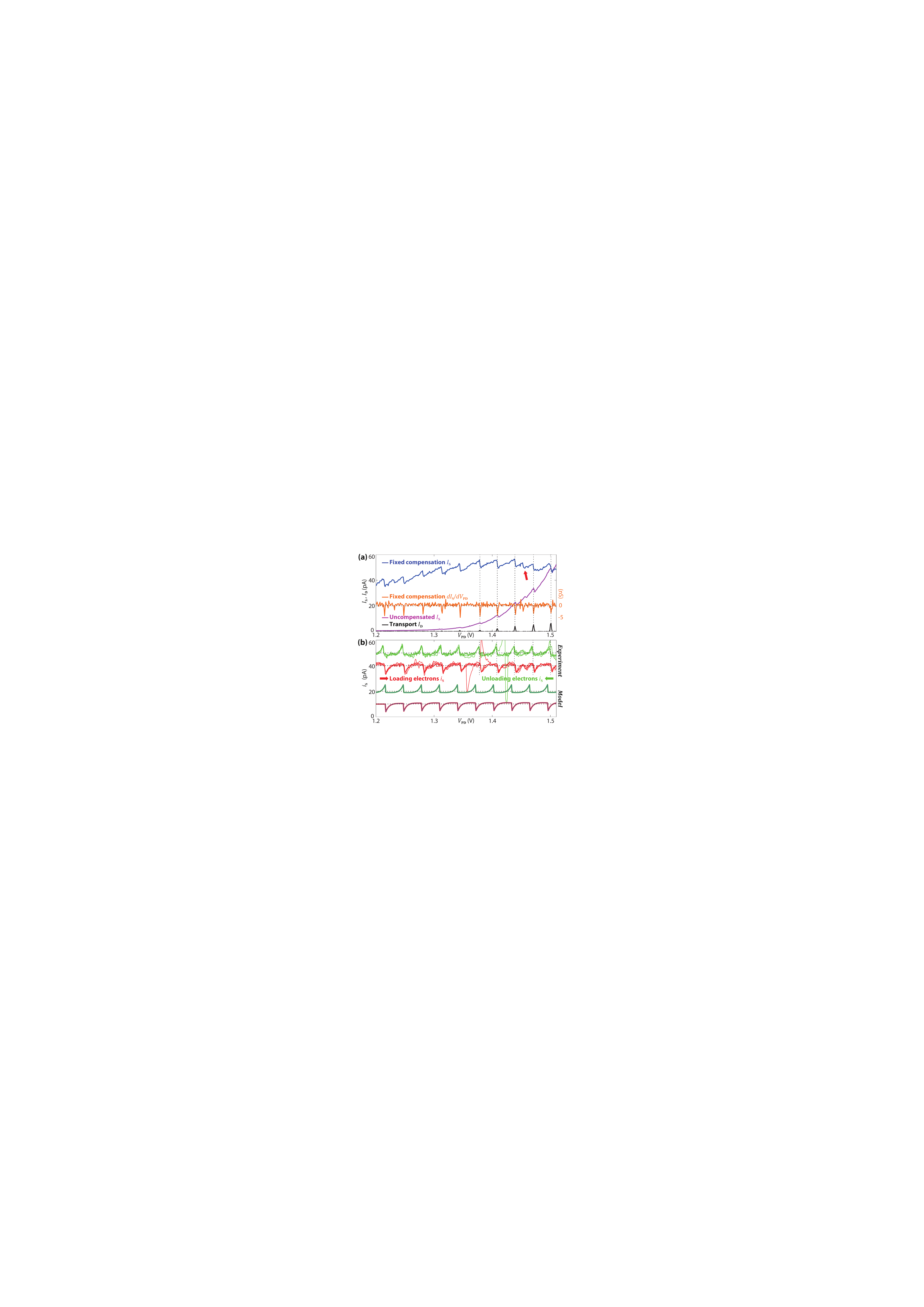}
\caption{(a) SET sensor current $I_\textrm{S}$ without compensation (magenta) and dot transport current $I_\textrm{D}$ (black).
		Fixed compensation is applied by linearly adjusting the sensor gate potential $V_\textrm{PS}$ and
		the compensated $I_\textrm{S}$ (dark blue) then operates within a fixed range,
		with a corresponding transconductance $dI_\textrm{S}$/$dV_\textrm{PD}$ (orange).
		(b) Sensor error current $i_\textrm{S}$ with dynamic feedback compensation applied
		while ramping $V_\textrm{PD}$ up (red) and down (green), with 5 overlaid traces.
		And modelled $i_\textrm{S}$ for the same range of experimental parameters.
		$I_\textrm{S}$ operating point is set to $I_\textrm{0}$ = 50~pA
		and the traces have been offset for clarity.
		Here, $\beta$ = 4~M$\Omega$ and $\gamma$ = 80~k$\Omega$.}
\label{fig2}
\end{figure}

Figure 2(a) shows the correlated signals of the quantum dot current
$I_\textrm{D}$ (black line) and the sensor current, $I_\textrm{S}$, with both
uncompensated (magenta) and fixed compensated (dark blue) sensor control.
When there is a single-electron transfer in the quantum dot (corresponding to a Coulomb peak in $I_\textrm{D}$),
an abrupt change appears in the sensor current trace $I_\textrm{S}$ (uncompensated).
As we reduce $V_\textrm{PD}$, $I_\textrm{S}$ will gradually shift away from the Coulomb peak edge
because the sensor island is capacitively coupled to the dot plunger.
Hence, this does not give a uniform single-electron transfer detection signal.

Fixed compensation improves the sensitivity of the charge sensing technique, leading to the dark blue trace in Fig. 2(a). Here, we adjust the sensor gate voltage by $\Delta V_\textrm{PS}$,
in proportion to $\Delta V_\textrm{PD}$ at a fixed ratio. The sensor current $I_\textrm{S}$ now operates within a fixed range,
keeping approximately the same position near the edge of the Coulomb peak.
The transconductance $dI_\textrm{S}$/$dV_\textrm{PD}$ of the sensor current, obtained numerically,
is plotted as an orange trace in Fig. 2(a).
Each time the occupancy of the quantum dot changes by one electron it produces a sharp negative peak in $dI_\textrm{S}$/$dV_\textrm{PD}$.
The sensor also detects an additional charge movement near $V_\textrm{PD}$ = 1.46~V,
whereas $I_\textrm{D}$ does not show any corresponding transport current through the dot.

We are able to distinguish the main (gate-defined) quantum dot from other unintentional (disorder-induced) dots or traps
by observing the difference in amplitudes and the positions of the charge sensing signals.
Arrow in Fig. 2(a) indicate the detection of a charge rearrangement outside the main dot.
The Coulomb peaks in $I_\textrm{D}$ have a regular pattern,
whereas the observed ``trap" charging signal is between two Coulomb peaks and is not visible in $I_\textrm{D}$.
The ability to distinguish the origin of charge movements makes SET or QPC sensing a valuable characterization tool
for the study of quantum dots and has motivated its use in many experiments discussed earlier~
\cite{Elzerman2003,Simmons2007,Elzerman2004,Morello2010,DiCarlo2004,Petta2004,Johnson2005,Cassidy2007,Simmons2009,Nordberg2009,Xiao2010,Hu2007,Biercuk2006,Wang2010,Podd2010}.

Despite its utility, fixed compensation of the charge sensor suffers from two significant problems:
(i) the effect of slow charge drifts; and (ii) sudden and random charge rearrangements in the environment
that cause significant changes in the sensor current and which cannot be compensated for.
Maintaining the stability of the charge sensor output over a long period of time
is therefore difficult without some form of feedback.
Here, dynamic feedback is used to control the charge sensor and correct for both of the fore-mentioned problems.
A computer-controlled second-order feedback algorithm adjusts the plunger gate PS of the charge sensor,
by taking $I_\textrm{S}$ as the feedback signal and retuning $V_\textrm{PS}$ for each sample $x$ being measured.

Figure 1(c) shows the block diagram for this dynamically-controlled charge sensor, which can be described by the following equations:

\begin{subequations}
\begin{align}
V_\textrm{PS}[x+1]&=V_\textrm{PS}[x]- \beta {i_\textrm{S}}[x]- \Delta V_\textrm{PD} A_\textrm{C}[x]\\
A_\textrm{C}[x+1]&=A_\textrm{C}[x]+ \frac{\gamma }{\Delta V_\textrm{PD}}  {i_\textrm{S}}[x]
\end{align}
\end{subequations}

Here, $\Delta V_\textrm{PD}$ is the dot-plunger-gate step size and $A_\textrm{C}=C_\textrm{PD}/C_\textrm{PS}$
is the mutual capacitance ratio between the dot-plunger PD to sensor island and the sensor-plunger PS to sensor island [see Fig. 1(d)].
The parameter $\beta$ controls the first-order feedback, which governs the decay rate of the error current,
$i_\textrm{S}=I_\textrm{S}-I_\textrm{0}$, where $I_\textrm{0}$ is the sensor ac operating point.
Note that when we apply fixed compensation we have $\beta$ = 0 and $A_\textrm{C}$ is constant.
In practice, $A_\textrm{C}$ can also experience sudden changes in the presence of charge rearrangements
and the parameter $\gamma$ governs the decay rate of $A_\textrm{C}$ back to its steady-state value after an upset.
Both $\beta$ and $\gamma$ are chosen to give a feedback system with good stability and a reasonable response time,
while ensuring that $I_\textrm{S}$ operates at the desired operating point $I_\textrm{0}$.

\begin{figure}[t]
\includegraphics[width=8.5cm]{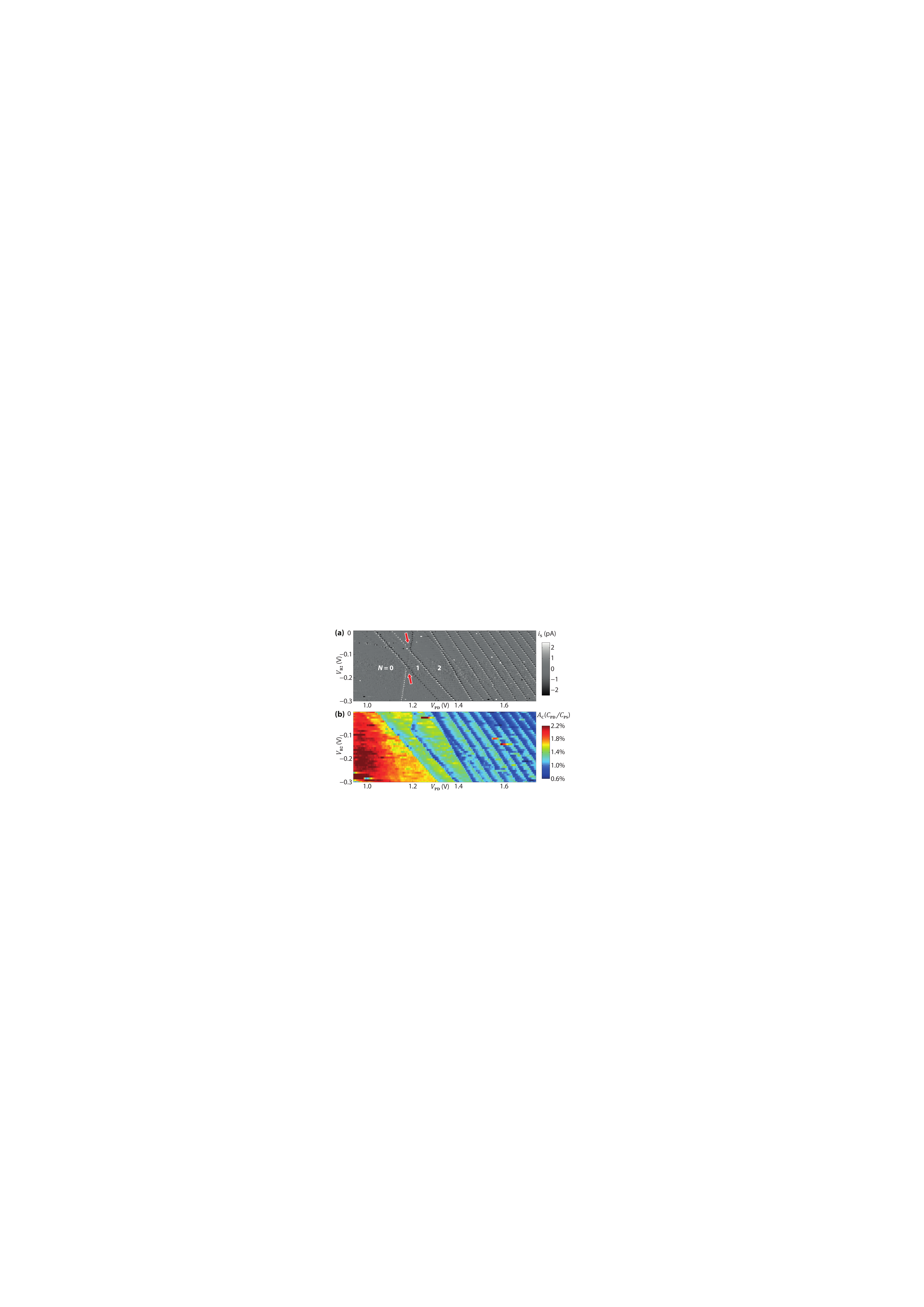}
\caption{(a) Stability diagram of $I_\textrm{S}$ showing the charge
transitions. $I_\textrm{S}$ now has an operating point $I_\textrm{0}$ of 80~pA.(b) Mutual capacitance ratio $A_\textrm{C}$
($C_\textrm{PD}/C_\textrm{PS}$), with positive sweeping only, showing
the change in $A_\textrm{C}$ when charge movement occurs.}
\label{fig3}
\end{figure}

Charge sensing results with dynamic feedback control are shown in Fig. 2(b).
Five overlaid traces for two opposing gate sweep directions are shown. When we increase the dot-plunger voltage $V_\textrm{PD}$(red traces),
the charge sensor detects electrons loading into the quantum dot,
and a sudden decrease in $i_\textrm{S}$ occurs for each loading event.
This drop triggers the feedback system to increase $V_\textrm{PS}$ in order to pull $I_\textrm{S}$ back to the operating point $I_\textrm{0}$.
In contrast, when we increase the dot-plunger voltage $V_\textrm{PD}$(green traces),
an electron moving out of the dot causes a sudden increase in $i_\textrm{S}$.
The feedback mechanism then gradually decreases $I_\textrm{S}$ back to $I_\textrm{0}$.
Figure 2(b) also shows the modeled values using the same parameters as the controller
and the measured physical properties of the device.
Good agreement between the model and the experimental results confirms that the feedback system is working as designed.

In the experimental data of Fig. 2(b) we observe a large upset event in the up-sweep (red trace) at $V_\textrm{PD}$ $\sim$ 1.35 V,
corresponding to what must have been a large and random charge rearrangement in the device.
Significantly, the feedback control enabled the sensor current $I_\textrm{S}$ to return to its optimal operating point $I_\textrm{0}$ after this event,
meaning that charge sensing could then continue.
With only fixed compensation applied, such a dramatic upset would most likely have shifted the sensor to a near zero current
and charge sensing would have been significantly impeded.
This demonstrates the utility of the dynamic feedback compensation for charge sensing over long time periods,
which is often necessary in order to fully characterize a quantum dot system or other nanostructures.

In order to further assess the robustness of the technique
we studied the charge state of the quantum dot by mapping
the sensor error current $i_\textrm{S}$ as a function of the dot plunger gate $V_\textrm{PD}$
and one of the dot barrier gates $V_\textrm{B2}$ -- see Fig. 3(a).
The interlaced positive (negative) sweeps of the plunger gate voltage produce dips (peaks) in $i_\textrm{S}$,
thus mapping the charge transitions in the quantum dot.
Loading of an electron is represented by a white pixel and unloading as black in Fig. 3(a).
Randomly occurring white and black pixels corresponds to charging and ionization of charge traps.
This data was obtained over 10 hours and the level of sensitivity
and sensor operating point remained constant over this entire period,
despite the occurrence of a number of upset events.

Several interesting features are observed in Fig. 3(a).
The main quantum dot contains no electron on the far left side of the plot, since there are no more charge transitions.
The first two transition lines have a different slope to the other regular ones. We believe that at low $V_\textrm{PD}$, the shape of the potential well in the quantum dot may have deformed due to the presence of local disorder. This changes the coupling of the dot-plunger gate to the quantum dot resulting in a different slope. Another feature observed is the hysteresis present when loading/unloading the first two electrons,
with the loading and unloading events occurring at different values of $V_\textrm{PD}$.
The red arrows show the intersections of the main transitions and a line where hysteresis occurs.
The origin of this hysteresis is not fully understood,
but is most likely related to non-equilibrium processes and coupling between the main dot and a nearby charge trap.

The mutual capacitance ratio $A_\textrm{C}$ calculated by the feedback control system
provides an additional parameter to aid understanding of the quantum dot system.
It represents the change in $C_\textrm{PD}$ with different electron occupancy,
assuming $C_\textrm{PS}$ stays constant. Figure 3(b) plots $A_\textrm{C}$, obtained simultaneously with the data in Fig. 3(a),
and supports the assignment $N$ = 0 for the electron occupancy at low gate voltages.
When $N$ = 0, $C_\textrm{PD}$ increases significantly,
due to the absence of the capacitive screening effect of the electrons in the dot.

We note that a similar SET architecture by Angus \textit{et al.}~\cite{Angus2008} demonstrated the capability of radio-frequency charge detection with sensitivities of better than 10 $\mu e/\sqrt{\textrm{Hz}}$. Hence, we expect that our SET structure and digital feedback system should be able to perform time-averaged sensing at MHz bandwidth. However in this experiment, the electrical setup is designated for low-frequency measurements only.

In conclusion, we have used a silicon single electron transistor
to demonstrate charge sensing of a nearby silicon MOS quantum dot.
We observed single electron occupancy of the dot and demonstrated the benefits of dynamic feedback control.
The control algorithm is highly robust against charge drift and random charge upset events,
enabling measurement stability for long periods (up to hours) and over a wide range of gate biases.
This device architecture and sensing technique has excellent potential for future experiments
such as single-shot electron spin readout and charge sensing in double quantum dots.

The authors thank D. Barber and B. Starrett for technical support and T. Duty for helpful discussions. This work was supported by the Australian Research Council (under contract CE110001027),
and by the U.S. National Security Agency and U.S. Army Research Office (under contract W911NF-08-1-0527).

\end{document}